\documentclass{rnaastex}

\usepackage{hyperref}
\hypersetup{pdfstartview={XYZ null null 1.00}}   

\begin{document}

\title{Triangulation Tracking of a Radially Propagating MHD Wave in the AIA 1600 Acoustic Power Maps in Active Region 12193}

\correspondingauthor{Teresa Monsue}
\email{teresa.monsue@vanderbilt.edu}

\author[0000-0003-3896-3059]{Teresa Monsue}
\affil{Department of Physics and Astronomy, Vanderbilt University, Nashville, TN 37235, USA}
\affiliation{NASA Goddard Space Flight Center, Greenbelt, MD 20771, USA}




\section{} 

\par
For decades it has been established that the amount of energy released by solar flares excites the acoustic oscillations propagating on the surface of the Sun \citep{Wolff1972}.  It is believed that these flares can excite velocity oscillations in active regions, especially those regions where a higher class solar flare has taken place \citep{Kumar2006}.  However, questions arise as to how the behaviors of acoustic oscillations within such a chaotic environment can birth other waves of the MHD type.  \textit{Can we observe such events?}
\vspace*{0.15in}
\par
\begin{figure}[ht!]
\begin{minipage}{8in}
  \centering
  \hspace*{-1.0in}
  \raisebox{-0.5\height}{\includegraphics[height=3.5in]{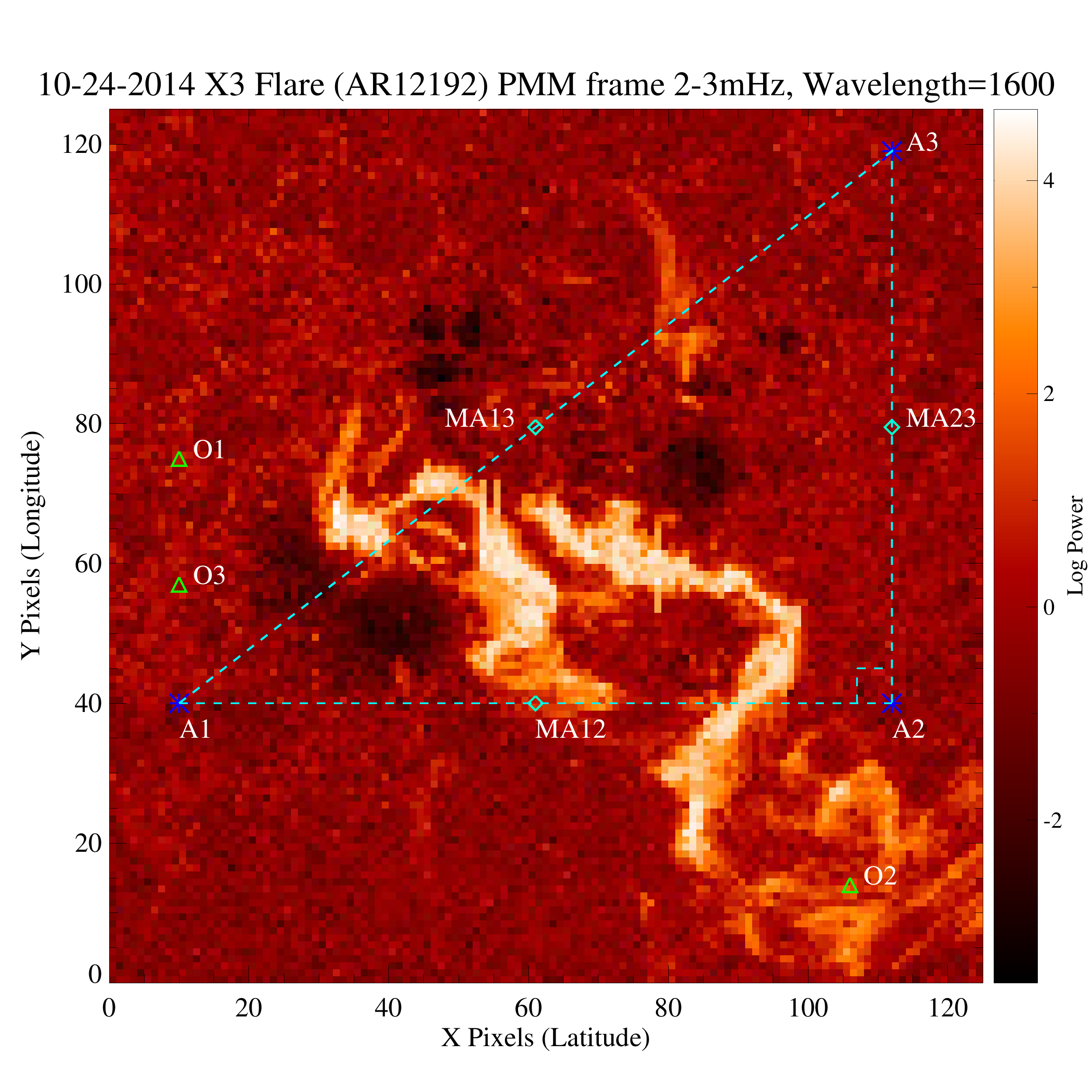}}
  \raisebox{-0.5\height}{\includegraphics[height=6.5in]{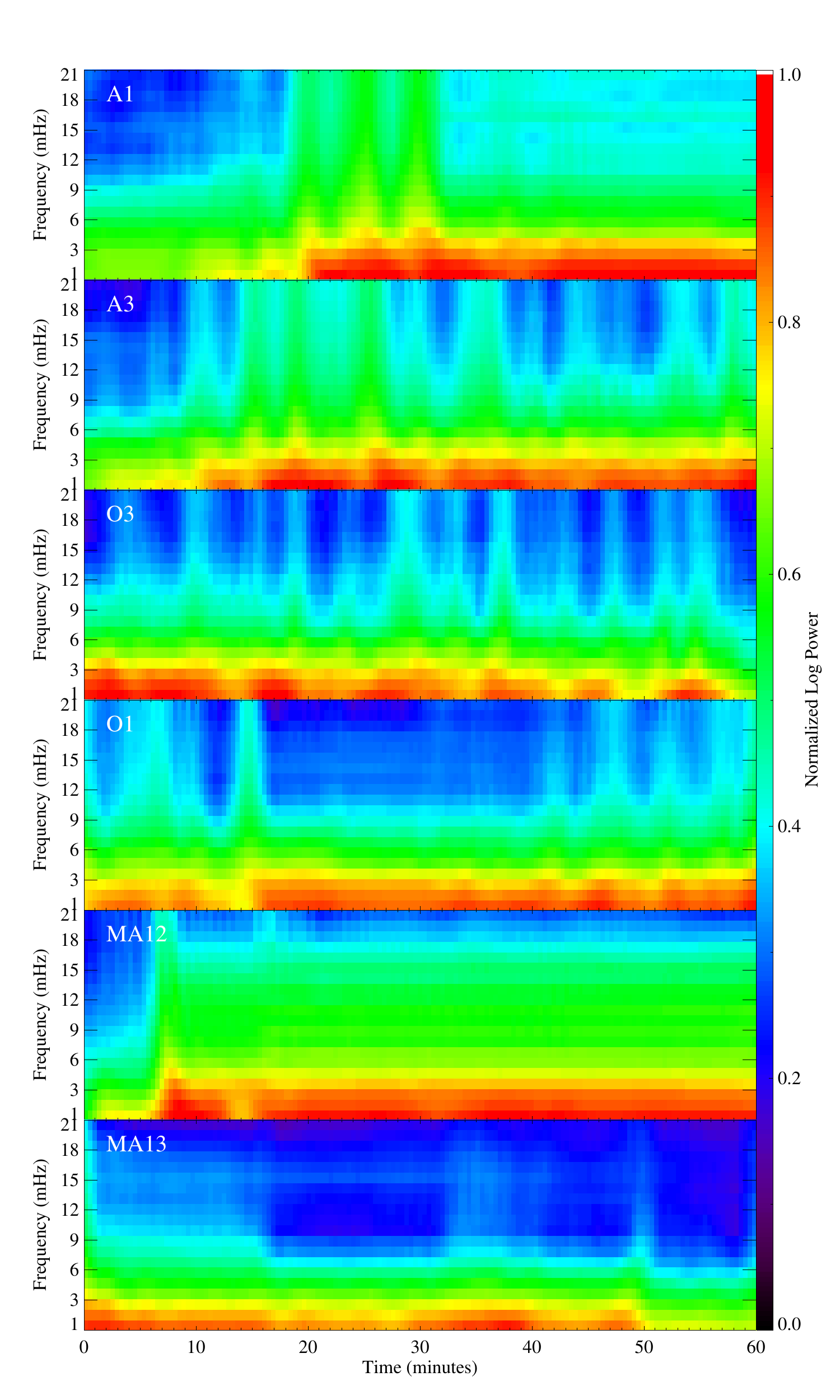}}
  \end{minipage}
\caption{\textit{(Left)}  Illustrated above is a PMM frame (the region in measured in power) in the 2--3mHz band, windowed at 30--90 minutes.  Sampled 90$^{\circ}$ triangular region in flaring region AR12192 (AIA1600 wavelength).  The large region of interest is $125\times 125$ pixels encompassing the active flare region.  \textit{(Right)} Spectrogram plots of corresponding regions, A1, A3, O3, O1, MA12 and MA13.\label{fig:f1}}
\end{figure}

We employed a method first devised by Jackiewicz \& Balasubramaniam (\citeyear{Jackiewicz2013}) and then studied further by Monsue \textit{et al.} (\citeyear{Monsue2016}) in which we observe an active region in the spatial-frequency domain as it evolved through time, therefore creating a power map movie (PMM).  We chose an X3-class solar flare that occurred on October 24, 2014 at 21:41 U.T., in active region NOAA AR12192.  Through PMMs, we were able to study the spatial and temporal frequency information per pixel within the AIA1600 wavelength images, from NASA's Solar Dynamics Observatory \citep{Pesnell2012}, by running the Fast Fourier Transform \textit{(FFT)} on each pixel.  Our observation window is two hours of images, surrounding the flare.  We set the window length to run the FFT to one hour as we move through the datacube of the 1600 images, enabelling us to study the broadband behavior of the acoustic oscillations within the frequency bands of 1--21mHz.  We reached a Nyquist frequency observation limit of ~21mHz, since the AIA1600 cadence is 24 seconds.  We also averaged the intensity for each pixel to 5-spatial pixels for each 1600 image frame before running the FFT.  We observed in the 2--4 mHz PMMs of the AIA1600 images a radially propagating enhancement occurring during the flaring event.  We suspected that this propagating enhancement is due to the creation of a MHD wave moving out away from the flare.  We studied this closer by devising a way to triangulate the movement of the MHD wave by taking a 90$^{\circ}$ triangular sampling to see how the pixels behaved by constructing a spectrogram at each sampled region (e.g. Figure \ref{fig:f1}).  For a sampling of spectrograms of some of the regions (A1, A3, O3, O1), we were able to see oscillatory behavior.  However, we are unable to conclude if this oscillatory behavior is attributable to the MHD wave, or aliasing due the behavior of the FFT.  However, this is a work in progress.

\acknowledgments

T.M. acknowledges funding from a fellowship and a grant from NASA and the use of Vanderbilt's ACCRE cluster to complete this work, and the advice from Drs. Dean Pesnell (NASA Goddard) and Frank Hill (NSO).

\end{document}